\documentstyle[prl,aps,epsf,floats]{revtex}

\begin{document}         
\twocolumn[
\hsize\textwidth\columnwidth\hsize\csname@twocolumnfalse\endcsname

\title{Quasi-one-dimensional superconductors: from weak to strong
magnetic field }

\author{N. Dupuis}
\address{Department of Physics, University
of Maryland, College Park, MD 20742-4111, USA \\
Laboratoire de Physique des Solides,
Universit\'e Paris-Sud, 91405 Orsay, France}
\date{To be published in Journal of Superconductivity, Proceedings of
the Workshop on {\it Exactly Aligned Magnetic Field Effects in
Low-Dimensional Superconductors} (Kyoto, Nov. 1998) }
\maketitle

\begin{abstract}
We discuss the possible existence of a superconducting phase at high
magnetic field in organic quasi-one-dimensional conductors. We
consider in particular (i) the formation of a
Larkin-Ovchinnikov-Fulde-Ferrell state, (ii) the role of a 
temperature-induced dimensional crossover occuring when the
transverse coherence length $\xi_z(T)$ becomes of the order of the lattice
spacing, and (iii) the effect of a
magnetic-field-induced dimensional crossover resulting from the
localization of the wave functions at high magnetic field. In the case
of singlet spin pairing, only the combination of (i) and (iii) yields
a picture consistent with recent experiments in the Bechgaard salts
showing the existence of a high-field superconducting phase. 
We point out that the vortex lattice is expected to exhibit
unusual characteristics at high magnetic field.  
\end{abstract}
]

\section{INTRODUCTION}

According to conventional wisdom, superconductivity and high magnetic field are
incompatible. A magnetic field acting on the orbital electronic motion breaks 
down time-reversal symmetry and ultimately restores the metallic phase. In the 
case of singlet pairing, the coupling of the field to the electron spins 
also suppresses the superconducting order (Pauli or Clogston-Chandrasekhar 
limit [1]). 

Recently, this conventional point of view has been challenged, both 
theoretically [2-6] and experimentally [7], in particular in
quasi-1D organic materials.  
Because of their {\it open Fermi surface}, these superconductors exhibit unusual
properties in presence of a magnetic field. As first recognized by
Lebed' [2], the
possible existence of a superconducting phase at high magnetic field results
from a {\it magnetic-field-induced dimensional crossover} that freezes
the orbital mechanism of destruction of superconductivity. Moreover,
the Pauli pair breaking (PPB) effect can be largely compensated by the
formation of a Larkin-Ovchinnikov-Fulde-Ferrell (LOFF) state [2-5,8].

The aim of this paper is to discuss three aspects of high-field
superconductivity in quasi-1D conductors: the formation of a LOFF
state (i), and the respective roles of temperature-induced (ii) and
magnetic-field-induced (iii) dimensional crossovers. 

We consider a quasi-1D superconductor with an open Fermi
surface corresponding to the dispersion law ($\hbar=k_B=1$)
\begin{equation}
E_{\bf k}=v_F(|k_x|-k_F)+t_y\cos(k_yb)+t_z\cos(k_zd)+\mu.
\end{equation}
$\mu$ is the Fermi energy, and $v_F$ the Fermi velocity for the motion
along the chains. $t_y$ and $t_z$ ($t_y\gg t_z$) are the transfer
integrals between chains. The magnetic field $H$ is applied along the
$y$ direction and we denote by $T_{c0}$ the zero-field transition
temperature. [In Bechgaard salts, $T_{c0}\sim 1$ K, and $t_c=t_z/2$ is in
the range 2-10 K [3,4,7].]

\section{LOFF STATE IN QUASI-1D SUPERCONDUCTORS}

We first discuss the effect of a magnetic field acting on the electron
spins. Larkin and Ovchinnikov, and Fulde and Ferrell, have shown that the 
destructive influence of Pauli paramagnetism on superconductivity can be
partially compensated by pairing up and down spins with a non-zero total 
momentum [8]. At low temperature (i.e. high magnetic field), when $T\leq T_0
\simeq 0.56T_{c0}$, this non-uniform state becomes more stable than the 
uniform state corresponding to a vanishing momentum of Cooper pairs. 

Quasi-1D superconductors appear very particular with respect to the existence
of a LOFF state [4]. The fundamental reason is that, because of the quasi-1D 
structure of the Fermi surface, the partial compensation of the Pauli pair
breaking effect by a spatial modulation of the order parameter is much more 
efficient than in isotropic systems. Indeed, a proper choice of the
Cooper pairs momentum allows one to keep one half of the phase space available 
for pairing whatever the value of the magnetic field. Thus, the
critical field $H_{c2}(T)\propto 1/T$ diverges at low temperature [4].

\section{TEMPERATURE-INDUCED DIMENSIONAL CROSSOVER}

At low temperature, strongly anisotropic superconductors may exhibit a
dimensional crossover that allows the superconducting phase to persist
at arbitrary strong field (in the mean-field approximation) in the
absence of PPB effect. 

Within the Ginzburg-Landau theory, at high temperature ($T\sim
T_{c0}$) the mixed state is an (anisotropic) vortex lattice. The critical
field $H_{c2}(T)$ is determined by $H_{c2}(T)=\phi_0/2\pi\xi_x(T)\xi_z(T)$ 
where $\xi_x$ and $\xi_z$ are the superconducting coherence lengths
and $\phi_0$ the flux quantum. When the anisotropy is large enough,
i.e. when $t_z \stackrel{\textstyle <}{\sim } T_{c0}$, $\xi_z(T)$ can
become at low temperature of the order or even smaller than the lattice
spacing $d$. Vortex cores, which have an extension  
$\sim \xi_z(T)$ in the $z$ direction, can then fit between 
planes without destroying the superconducting order in the planes. The 
superconducting state is a Josephson vortex lattice and is always stable
at low temperature for arbitrary magnetic field (see Fig.~3 in
Ref.~4). A proper description of this situation, which takes 
into account the discreteness of the lattice in the $z$ direction, 
is given by the Lawrence-Doniach model [4,9].

It is tempting to conclude that this temperature-induced dimensional
crossover, together with the formation of a LOFF state, could lead to
a diverging critical field $H_{c2}(T)$ at low temperature. It has been shown in 
Ref.~4 that this is not the case: the PPB effect strongly suppresses
the high-field superconducting phase (this point is further discussed
in sect.~4.1 (see Fig.~1.b)). Therefore the
temperature-induced dimensional crossover cannot explain the critical
field $H_{c2}(T)$ measured in Bechgaard salts [7] in the case of spin
singlet pairing.

\section{MAGNETIC-FIELD-INDUCED DIMENSIONAL CROSSOVER}
 
The microscopic justification of the Ginzburg-Landau or Lawrence-Doniach 
theory of the mixed state of type II superconductors is based on a 
semiclassical approximation (known as the semiclassical phase integral 
or eikonal approximation) that completely neglects the quantum effects of
the magnetic field. At low temperature (or high magnetic field) and in
sufficiently clean superconductors, when $\omega_c\gg T,\tau$ 
($\omega_c=eHdv_F/\hbar$ being the frequency of the semiclassical
electronic motion, and $\tau$ the elastic scattering time), these
effects cannot be neglected and an exact description  of the field is
required.  

To be more specific, we write the Green's function (or electron
propagator) as [10]
\begin{equation}
G({\bf r}_1,{\bf r}_2)= \exp \Bigl \lbrace ie\int _{{\bf r}_1}^{{\bf r}_2}
d{\bf l}\cdot {\bf A} \Bigr \rbrace \bar G({\bf r}_1-{\bf r}_2),
\end{equation}
where $\bf A$ is the vector potential. The Ginzburg-Landau or Lawrence-Doniach 
theory identifies $\bar G$ with the Green's function $G_0$ in the absence of
magnetic field. The latter intervenes only through the phase factor 
$ie\int _{{\bf r}_1}^{{\bf r}_2}d{\bf l}\cdot {\bf A}$, which breaks down 
time-reversal symmetry and tends to suppress the superconducting order.

When $\omega_c\gg T,1/\tau$, the approximation $\bar G=G_0$ 
breaks down and a proper treatment of the field is required. In isotropic
systems, $\bar G$ includes all the information about Landau level
quantization. In strongly anisotropic conductors, it describes a 
magnetic-field-induced dimensional crossover [2,3], i.e. a confinement of the
electrons in the planes of highest conductivity. The same conclusion can
be reached by considering the semiclassical equation of motion $\hbar d{\bf k}
/dt=e{\bf v}\times {\bf H}$ with ${\bf v}=\mbox{\boldmath$\nabla$}
E_{\bf k}$. The corresponding electronic orbits in real space are of the form 
(neglecting for simplicity the (free) motion along the field)
$z=z_0+d(t_z/\omega_c)\cos(\omega_c x/v_F)$.
The electronic motion is extended along the chains (and the magnetic field
direction), but confined with respect to the $z$ direction 
with an extension $\sim d(t_z/\omega_c)\propto 1/H$. In very
strong field $\omega_c\gg t_z$, the amplitude of the orbits becomes smaller 
than the lattice spacing $d$ showing that the electronic motion is localized in
the $(x,y)$ planes. The latter being parallel to the magnetic field, the
orbital frustration of the superconducting order parameter vanishes (there is
no magnetic flux through the 2D Cooper pairs located in the $(x,y)$ planes).

\subsection{Large anisotropy}

Fig.~1.a shows the phase diagram in the exact mean-field approximation
in the case of a weak interplane transfer $t_z/T_{c0}\simeq 1.33$. $Q$
is a pseudo momentum for the Cooper pairs in the field [3-5]. At
high temperature ($T\sim T_{c0}$), the mixed state is an Abrikosov
vortex lattice, and $T_c$ decreases linearly with the field. $T_c$
does not depend on $Q$ in this regime, which is shown symbolically by
the shaded triangle in Fig.~1.a. For
$H\sim 0.3$ T, the system undergoes a temperature-induced dimensional
crossover, which leads to an upward curvature of the transition
line. This dimensional crossover selects the value $Q=0$ of the pseudo
momentum. The PPB effect becomes important for $H\sim 2$ T and
leads to a formation of a LOFF state ($Q$ then switches to a finite
value). At higher field, $Q\simeq 2\mu_BH/v_F$ ($\mu_B$ is the Bohr
magneton), and $T_c$ exhibits again an upward curvature. 

Fig.~1.b shows the phase diagram obtained in the Lawrence-Doniach
model [4]. The metallic phase is restored above a field $H\sim 2.7$ T,
and the LOFF state is stable only in a narrow window around $H\simeq
2.6$ T. Only when the magnetic-field-induced dimensional 
crossover is taken into account does the LOFF state remain stable at very
high magnetic field. In a microscopic picture, the dimensional crossover shows
up as a localization of the single-particle wave functions with a concomitant
quantization of the spectrum into a Wannier-Stark ladder (i.e. a set of 1D
spectra if we neglect the energy dispersion along the field). This is
precisely this 
quantization that allows one to construct a LOFF state in a way similar to
the 1D or (zero-field) quasi-1D case [5]. Thus, when the field is treated 
semiclassically, the region of stability of the LOFF state in the $H-T$ plane
becomes very narrow as in isotropic 2D or 3D systems [8,11]. 

\begin{figure}[h]
\epsfysize 9.cm 
\epsffile[145 160 505 650]{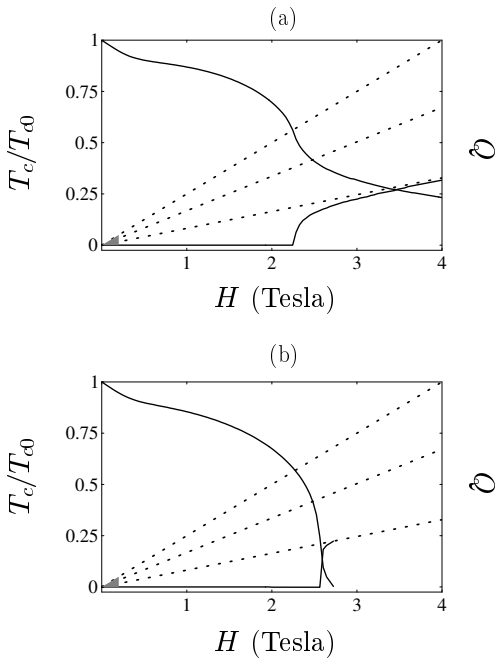}
%\epsfysize 14.cm 
%\epsffile[145 -50 505 650]{fig1.ps}
\caption{(a) Phase diagram for $t_z/T_{c0}\simeq 1.33$. $Q$ is a
pseudo momentum for the Cooper pairs in the field. The three doted
lines correspond to $Q=2\mu_B/v_F$, $G-2\mu_B/v_F$, $G$ ($G=\omega_c/v_F$). 
(b) Phase diagram in the Lawrence-Doniach model. }
\label{Fig1}
\end{figure}
\vspace{1cm}

\subsection{Smaller anisotropy}

For a smaller anisotropy, the coherence length 
$\xi_z(T)$ is always larger than the spacing between chains: $\xi_z(T)\geq 
\xi_z(T=0)>d$. There is no possibility of a temperature-induced dimensional
crossover. 

Fig.~2.a shows the phase diagram without the PPB effect for
$t_z/T_{c0}\simeq 4$. The low-field Ginzburg-Landau regime
(corresponding to the shaded triangle in Fig.~2) is followed by a
cascade of superconducting phases separated by first-order
transitions. These phases correspond to either $Q=0$ or
$Q=G\equiv\omega_c/v_F$ [3]. In the quantum regime, the field-induced
localization of the wave functions plays a crucial role in the pairing
mechanism. The transverse periodicity $a_z$ of the vortex lattice is
not determined by the Ginzburg-Landau coherence length $\xi_z(T)$ but
by the magnetic length $d(t_z/\omega_c)$. The first order phase
transitions are due to commensurability effects between the crystal
lattice spacing $d$ and $a_z$: each phase corresponds to a periodicity
$a_z=Nd$ ($N$ integer). $N$ decreases by one unit at each phase
transition. The mixed state evolves from a triangular
Abrikosov vortex lattice in weak field to a triangular Josephson
vortex lattice in very high field (where $N=2$). It has been pointed
out that $a_z$ decreases in both the Ginzburg-Landau and quantum
regimes, but {\it increases} at the crossover between the two regimes
where $\omega_c\sim T$ [3]. This suggests that the mixed state exhibits
unusual characteristics in the quantum regime. Indeed, the amplitude
of the order parameter and the current distribution show a symmetry of
laminar type. In particular, each chain carries a non-zero total
current (except the last phase $N=2$). We expect these unusual
characteristics to influence various physical measurements. 

Fig.~2.b shows the phase diagram when the PPB effect is taken into
account. There is an interplay between the cascade of phases and the
formation of a LOFF state. The latter corresponds to phases with
$Q=2\mu_BH/v_F$ and $Q=G-2\mu_BH/v_F$.

\begin{figure}[h]
\epsfysize 9.cm 
\epsffile[145 160 505 650]{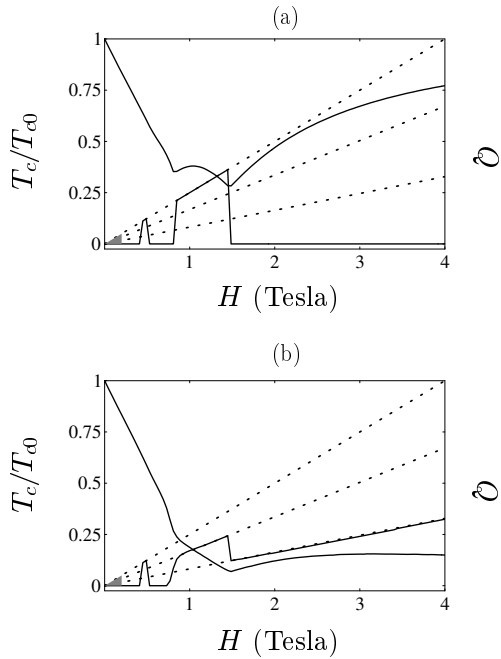}
%\epsfysize 14.cm 
%\epsffile[145 -50 505 650]{fig2.ps}
\caption{Phase diagram for $t_z/T_{c0}\simeq 4$ without (a) and with
(b) PPB effect. }  
\label{Fig2}
\end{figure}

\section{CONCLUSION}

Our discussion shows that a temperature-induced dimensional
crossover could explain recent experiments in the Bechgaard salts only
in the case of triplet pairing, although this would require  a value
of the interchain coupling $t_z$ slightly smaller than what is
commonly expected [3,4,7]. In the more likely case of singlet pairing,
the existence of a high-field superconducting phase in quasi-1D
conductors may result from a magnetic-field-induced dimensional
crossover and the formation of a LOFF state. The crossover between the
semiclassical Ginzburg-Landau and quantum regimes, which occurs when
$\omega_c\sim T$, is accompanied by an increase of the transverse
periodicity of the vortex lattice. This, as well as the
characteristics of the vortex lattice in the quantum regime (laminar
symmetry of the order parameter amplitude and the current distribution),
suggests that the high-field superconducting phase in quasi-1D
conductors should exhibit unique properties.

\section{REFERENCES}
 
\noindent
1. A.M. Clogston, {\it Phys. Rev. Lett.} {\bf 9}, 266 (1962);
B.S. Chandrasekhar,  {\it Appl. Phys. Lett.} {\bf 1}, 7 (1962).  \\
2. A.G. Lebed',  {\it JETP Lett.} {\bf 44}, 114 (1986);
L.I. Burlachkov, L.P. Gor'kov and A.G. Lebed', {\it EuroPhys. Lett.}
{\bf 4}, 941 (1987). \\
3. N. Dupuis, G. Montambaux and C.A.R. S\'a de Melo, {\it Phys.
Rev. Lett.} {\bf 70}, 2613 (1993); N. Dupuis and G. Montambaux,
{\it Phys. Rev. B} {\bf 49}, 8993 (1994). \\
4. N. Dupuis, {\it Phys. Rev. B} {\bf 51} 9074 (1995). \\
5. N. Dupuis, {\it Phys. Rev. B} {\bf 50}, 9607 (1994); {\it
J. Phys. I France} {\bf 5} 1577 (1995). \\  
6. Y. Hasegawa and M. Miyazaki, J. Phys. Soc. Jpn {\bf 65}, 1028
(1996); M. Miyazaki and Y. Hasegawa, J. Phys. Soc. Jpn {\bf 65}, 3283
(1996). \\
7. I.J. Lee, A.P. Hope, M.J. Leone, and M.J. Naughton,
{\it Synth. Met.} {\bf 70}, 747 (1995); {\it Appl. Supercond.} {\bf 2}, 753
(1994); I.J. Lee, M.J. Naughton, G.M. Danner, and P.M. Chaikin,
{\it Phys. Rev. Lett.} {\bf 78}, 3555 (1997). \\
8. A.I. Larkin and Yu.N. Ovchinnikov, {\it Sov. Phys. JETP} {\bf 20}, 762
(1965); P. Fulde and R.A. Ferrell, {\it Phys. Rev.} {\bf 135}, A550 (1964). \\
9. W.E. Lawrence and S. Doniach, in {\it Proceedings of the 12th
International Conference on Low Temperature Physics LT12, Kyoto},
edited by E. Kanada (Academic, New York, 1970). \\
10. L.W. Gruenberg and L. Gunther, {\it Phys. Rev.} {\bf 176}, 606 (1968). \\ 
11.  L.W. Gruenberg and L. Gunther, {\it Phys. Rev. Lett.} {\bf 16}, 996
(1966).

\end{document}